# **Title**

An Updated Review of Methods and Advancements in Microvascular Blood Flow Imaging

Authors:

**Cerine Lal[1], Martin J Leahy[1, 2]**

[1]Tissue Optics and Microcirculation Imaging

Department of Applied Physics

National University of Ireland

Galway

[2]Adjunct Professor,

Royal college of surgeons in Ireland

Dublin

Correspondence:

Prof. Martin J Leahy

martin.leahy@nuigalway.ie

Tissue Optics and Microcirculation Imaging

Department of Applied Physics

National University of Ireland

Galway

# Abstract

There has been a consistent growth in research involving imaging of microvasculature over the past few decades. By 2008, publications mentioning the microcirculation had grown more than 2000 per annum. Many techniques have been demonstrated for measurement of the microcirculation ranging from the earliest invasive techniques to the present high speed, high resolution non-invasive imaging techniques. Understanding the microvasculature is vital in tackling fundamental research questions as well as to understand effects of disease progression on the physiological wellbeing of an individual. We have previously provided a wide ranging review [38] covering most of the available techniques and their applications. In this review, we discuss the recent advances made and applications in the field of microcirculation imaging.

# List of Abbreviations

SNRs - signal to noise ratios

CT – Computed tomography

PSF – Point spread function

CBF – Cerebral blood flow

SMC - Smooth muscle cells

OCT - Optical coherence tomography

$PO_2$ - Partial pressure of oxygen

µCBF - microvascular cerebral blood flow

STED - Stimulated emission depletion microscopy

SSADA - Split spectrum amplitude decorrelation angiography

SV - Speckle variance

OMAG - Optical microangiography

UHS-OMAG - Ultrahigh sensitive OMAG

mOMAG – Binary mask OMAG

SC - Stratum corneum

VE - Viable epidermis

PD - Papillary dermis

RD - Reticular dermis

DOAMG – Doppler OMAG

RBC – Red blood cells

IS/OS – Inner segment/ outer segment

RPE – Retinal pigment epithelium

$sO_2$ - Haemoglobin oxygen saturation

STFT - Short-time inverse Fourier transform

DLS – OCT – Dynamic light scattering OCT

IM - Intracellular motility

HSI - Hyperspectral imaging

LSCI - Laser speckle contrast

LDPI - Laser Doppler perfusion imaging

SDFI - Sidestream dark field imaging

IDFI – Incident dark field imaging

CCD – Charge coupled device

LED – Light emitting diode

DCS – Diffuse correlation spectroscopy

NIR – Near infrared light

$CMRO_2$ - Cerebral metabolic rate for oxygen

fNIRS - Functional near infrared spectroscopy

CWS - Continuous wave spectroscopy

CBV – Cerebral blood flow volume

rCBF - relative CBF

$rCMRO_2$ - relative $CMRO_2$

PAI - Photoacoustic imaging

# 1. Introduction

Microcirculation is the movement of blood through the circulatory system of an organism comprising of arterioles, capillaries, venules and lymphatics. The microvasculature is considered to begin where the smallest arteries deliver blood to the capillaries and to end where the blood is collected in the venules. Typically, the term 'micro' refers to objects (diameters in this case) smaller than 100 µm, so perhaps the simplest definition is that it includes all vessels less than 100 µm. The primary functions of the microcirculation include the regulation of blood pressure and tissue perfusion, regulation of vascular resistance, maintaining hydrostatic pressure at the capillary level, transfer of metabolites and gases and thermoregulation. The vast majority of exchange and interaction with the cells of the body occurs at the microcirculatory level where the ~17,000 km of vessels pass within 65 µm (figure 1(a)) of every healthy living cell [13]. The need for living cells to exist in close proximity to the microcirculation is exquisitely displayed in the corrosion castings of the human fingertip recently reported by Flahavan as shown in figure 1(b-d) [50].

Over the years there has been a tremendous increase in microcirculation based research owing to the significant role it plays in maintaining the healthy physiological state of an organ and also in addressing the fundamental research questions pertaining to angiogenesis, neovascularization, neurovascular coupling and in the effectiveness of drugs for a given treatment. Figure 2(a) shows the percentage increase in number of research articles published based on microcirculation over the past years. There has been a steady increase in microcirculation based research output over the years with a high publication rate in the field of medical sciences. One can see the impact of the availability of video recording technology in the late 1960s and laser Doppler based equipment from the 1980s. We can also see that the technology development, indicated by publications in the physical science journals experiencing a surge since 2008. This is likely to reflect rapid developments of 3D microcirculation imaging technology. Figure 2(b) shows the number of research articles published based on the search term '*microcirculation imaging*'.

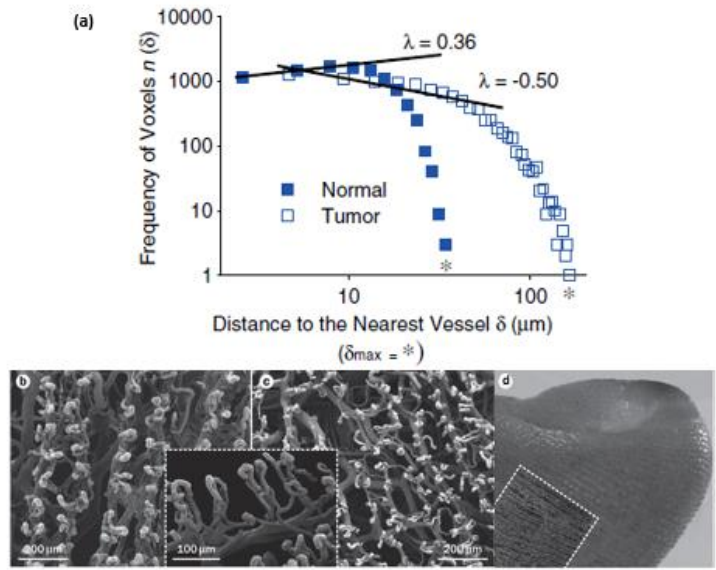

**Fig 1** (a) Graph showing the frequency of living cells as a function of distance from the nearest microcirculatory blood vessel, λ - measure of the shape of the spaces between vessels [13]. (b-d) Corrosion casting of the human fingertip reported by Flahavan showing (b) capillary loops on the palmar surface that follow the finger ridge pattern (c) capillary loops on the dorsal surface of the finger (d) Scanning electron micrograph of corrosion cast from a human finger (inset) superimposed onto the photograph of a human finger- tip showing the ridges [50].

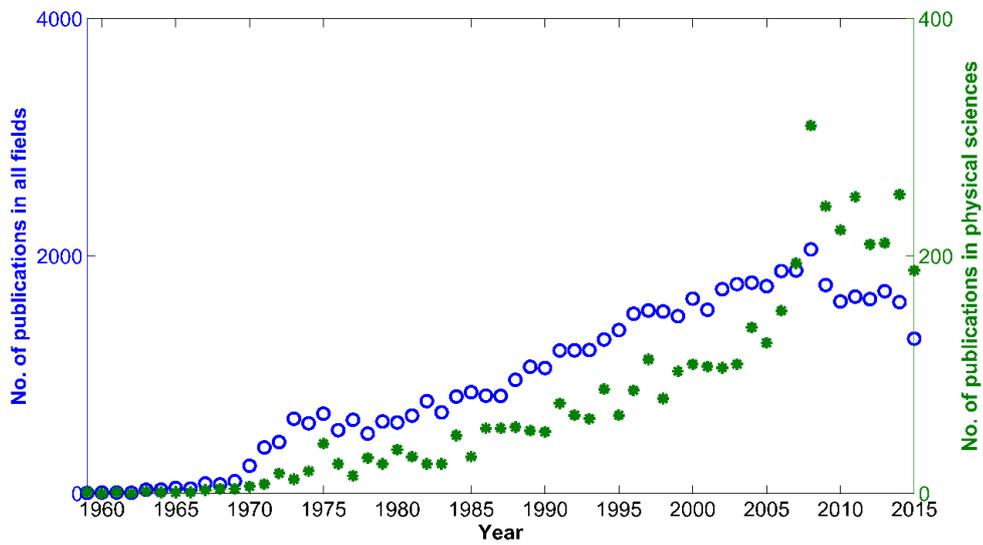

(a)

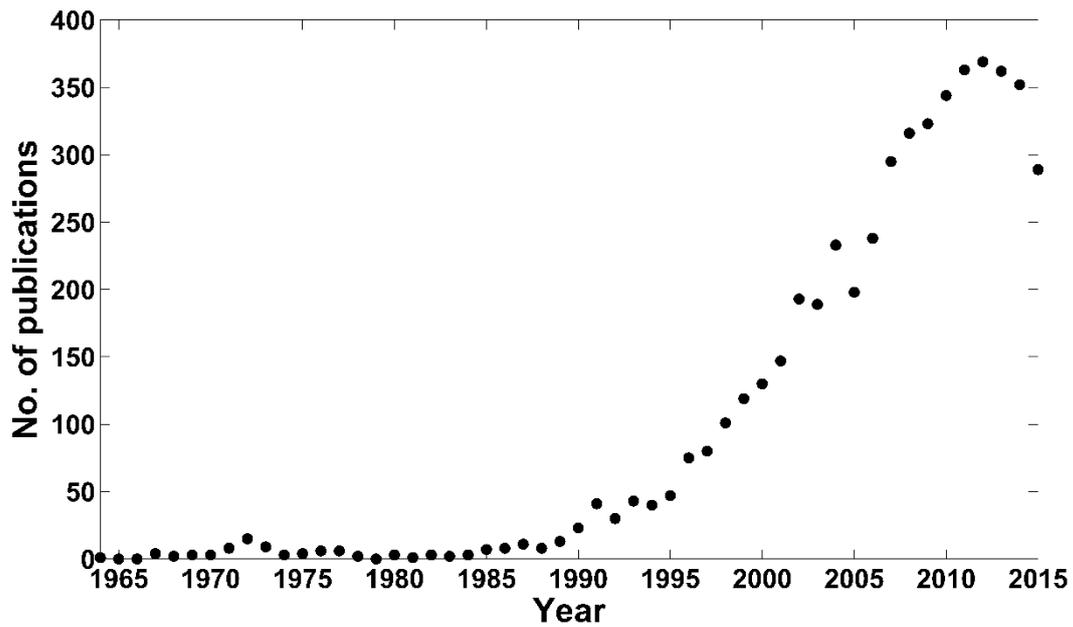

(b)

**Fig 2** (a) Graph showing the percentage increase in number of publications over the years using the search term 'microcirculation' (Source: Scopus) Blue axis: showing publications for all scientific fields Green axis: publications in the field of physical sciences. (b) Graph depicting the number of papers published using the search term 'microcirculation imaging' (Source: Scopus).

Currently available techniques for measurement of the microcirculation can be broadly divided into invasive and non-invasive modalities [5, 38]. Within the non-invasive imaging techniques, sound and light based imaging techniques are able to provide high resolution, clinically relevant information without disturbing the sample. Figure 3(a) shows the major non-invasive imaging modalities in terms of resolution and sampling depth.

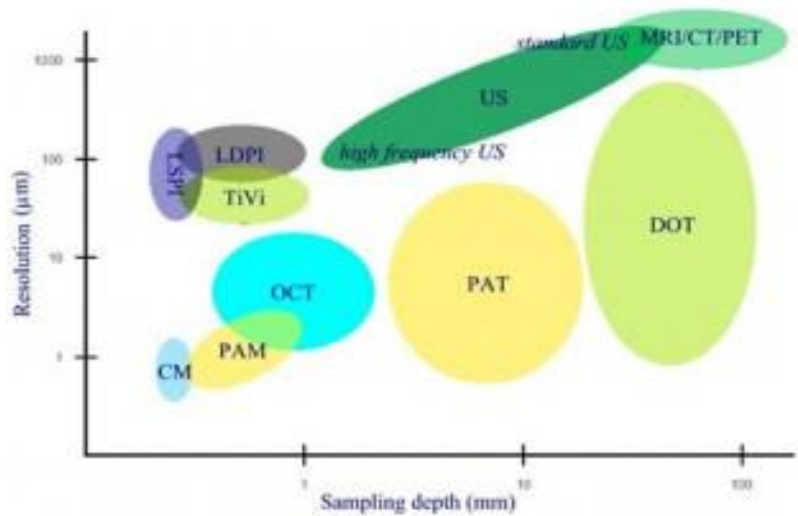

(a)

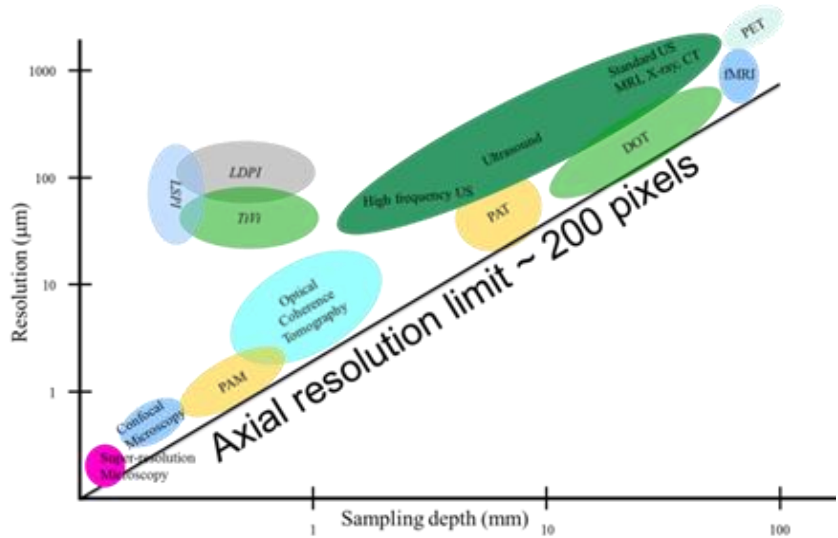

(b)

**Fig 3**. Illustration of various imaging modalities in terms of resolutions vs depth. (a) The distribution of imaging techniques as we used to discuss them. (b) The distribution of imaging techniques indicating an inverse relationship between imaging depth and voxel size.

Figure 3(b) indicates that for all imaging modalities, the depth of penetration within the tissue is inversely proportional to the voxel size in a 3-D volume. This is due to the attenuation of light/sound waves coming back from deeper regions within the sample thereby increasing the voxel size to collect sufficient detectable signals from deeper regions. This is likely due to the fact that signal to noise ratios (SNRs) are intrinsically proportional to the voxel volume or cube of spatial resolution (assuming its isotropic). This has been well established in field's such as XRay CT and ultrasound [51, 112, 9, 19]. In a noise limited scenario, the point spread function (PSF) which is described as the response of an imaging system to a point source input is dependent on its SNR. The measured PSF is related to the standard deviation (uncertainty) of $n$ measurements and this is proportional to $\sqrt{n}$ and hence local fluence ($F$) $\Delta V$, where $\Delta V$ is the resolved volume [12].

Daly et al. [38] described the various invasive and non-invasive techniques of microcirculation measurement and their applications in detail. This paper is intended to give an updated review of recent developments in the field of microcirculation imaging using optical techniques. The first section provides an updated review of microcirculation imaging using optical techniques. The optical imaging modalities for measurement of the microcirculation can be broadly divided into superficial imaging techniques, sub-surface imaging techniques and deep tissue imaging techniques. Superficial imaging techniques include the capillaroscopy, confocal microscopy, two photon imaging and stimulated emission depletion microscopy. The sub-surface imaging techniques consist of optical coherence tomography, hyperspectral imaging, side stream dark field imaging and incident dark field imaging. The deep tissue imaging techniques described include diffusion correlation spectroscopy, functional near infrared spectroscopy and photoacoustic tomography. Some of these imaging techniques which has not been covered in the earlier review [38] have been described in detail. The second section describes the recent applications of optical imaging techniques in understanding cerebral haemodynamics.

## 2 Review of microcirculation imaging methods

### 2.1 Superficial imaging techniques

### 2.1.1 Video capillaroscopy

Video capillaroscopy is a simple and inexpensive imaging technique mainly used to study the morphology of superficial capillary structures within the first few micro meters from the skin surface. This technique uses white /green light source to illuminate the tissue of interest through a high numerical aperture objective lens. Light backscattered from the tissue is collected by the objective lens and an image is formed on the CCD camera. Illuminating the tissue with green light which is the isobestic point of the absorption spectra of deoxy- and oxy-haemoglobin instead of conventional white light results in images with better sensitivity [6]. A recent study has confirmed this [144]. Recent capillaroscopy instruments such as the Capiscope (KK Technology) are equipped with epi-illumination that produces high contrast images with reduced blood cell blurring. Video capillaroscopy has been used to study micro-angiopathologies occurring in the nailfold in many diseases such as Raynaud's phenomenon, systemic sclerosis, scleroderma and connective tissue diseases [35, 59]. Figure 4 shows the schematic of the set up and typical image obtained from the instrument.

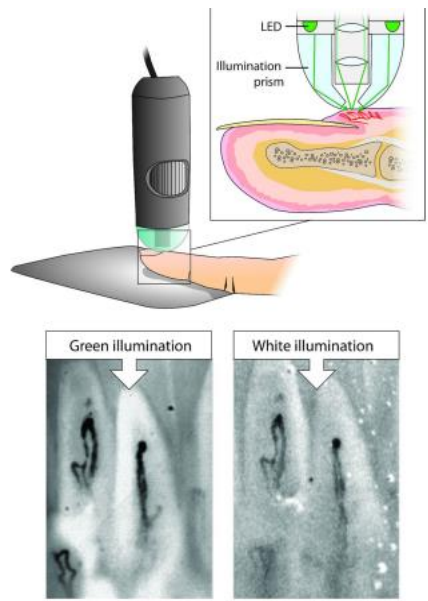

**Fig. 4** (top) Schematic of video capillaroscopy set up, (bottom) Capillary images obtained from green light and white light illumination respectively [144].

*2.1.2   Confocal microscopy and two photon imaging*

Confocal microscopy is a high resolution, high contrast 3-D imaging technique that can image upto a depth of 200 µm *in vivo* mainly from the epidermis and superficial layer of dermis [145]. The principle of confocal imaging was developed by Marvin Minsky in 1955 [145]. In this technique, a small spot within the tissue is illuminated by a highly focussed excitation beam. However, unlike conventional microscopy the reflected light from the focal spot within the tissue is projected through a pinhole aperture onto a light detector. The use of pinhole aperture allows only the focused light from the spot to pass through and completely eliminates the scattered light. Confocal microscopy uses reflected light or fluorescent light from the tissue for imaging. In order to create a 2-D image, the focal spot is scanned across the surface of the sample [38, 145]. Fluorescence confocal microscopy uses fluorescent dyes of fluorophores within the sample that fluoresce when stimulated by the excitation light. Fluorescence confocal microscopy has better sensitivity and specificity compared to reflection confocal microscopy. Figure 5(a) shows the schematic of confocal microscope.

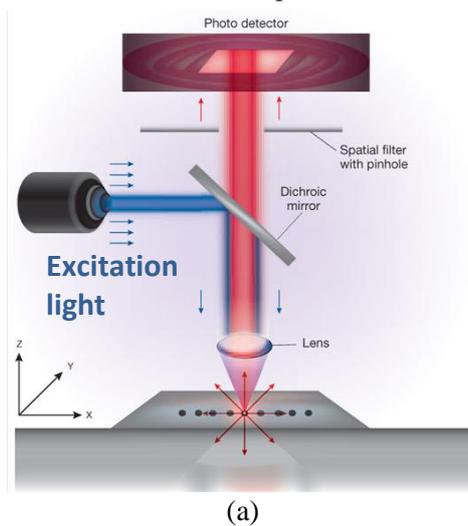

(a)

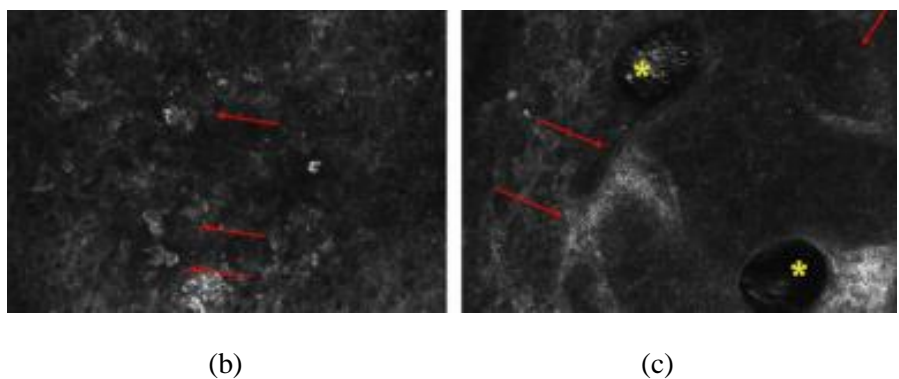

(b)                                       (c)

**Fig 5** (a) Schematic of confocal microscopy imaging [145]. (b) Reflectance confocal images of a dermal melanoma lesion showing the presence of several pagetoid cells (arrows) (c) Presence of cerebriform nests (arrows) and increased vascularization (asterisks) [109].

Confocal microscopy has been used to study the principles of vessel regression [53], erythrocyte properties [116], detecting skin lesions in oncology [109] and in studying the effect of glutamic acid on the blood- brain barrier and the neurovascular unit in glutaric acidemia I [70]. It has also been used to quantify capillary cell blood flow [30] and in evaluating cutaneous microcirculation and dermal changes in systemic sclerosis [135]. Figure 5(b) and 5(c) shows the reflectance confocal images of a dermal melanoma lesion showing increased vascularization.

Two photon microscopy invented by Denk and co-workers provides an imaging depth of 1 mm *in vivo* in tissues. It is based on the simultaneous absorption of two infrared photons by a fluorophore causing it to fluoresce. Tight focussing brought about by near-simultaneous absorption of photons causes the out-of-focus fluorescent light to be eliminated without the use of a pin hole as used in confocal microscopy. As the probability of near-simultaneous absorption is low, this technique requires the use of femtosecond lasers with short pulse duration for excitation [107]. Since it uses near infrared light, it has much deeper penetration within the tissue compared to confocal microscopy. Schematic of two photon microscopy is shown in figure 6.

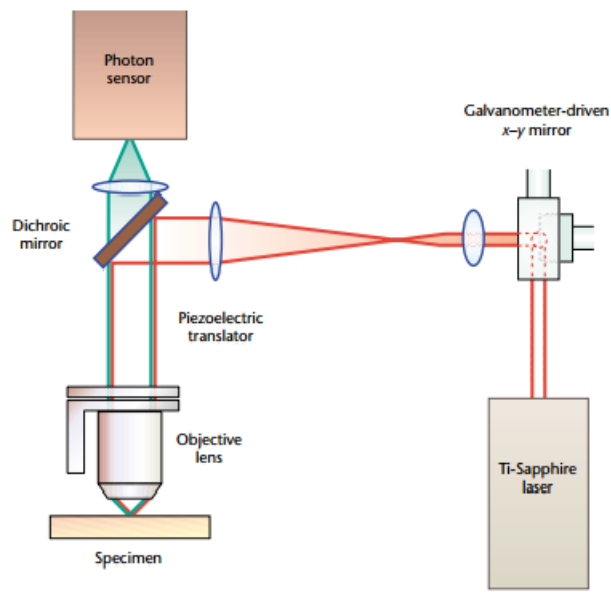

**Fig 6** Schematic of two photon microscopy system [107].

Two photon microscopy is used in rodent cerebrovascular analysis and neuro-coupling studies [37, 65, 47, 60, 61], cancer research [95], microvasculature [69, 140] and drug delivery [80]. A recent study by Sakadžić et al. [115] addresses the question of how the organization and morphology of cerebral microvasculature can provide for adequate tissue oxygenation during different metabolic activity levels. To address this challenge, they developed a multi-modal microscopy imaging setup based on Two-Photon Microscopy and Doppler-optical coherence tomography (OCT) to map partial pressure of oxygen ($PO_2$) and CBF respectively. Figure 7(a) shows the schematic of their experimental set up. They measured $PO_2$ distribution in the microvascular segments of primary somatosensory cortex down to a cortical depth of 450 µm through a sealed cranial window in mice models kept under normoxic normocapnic and mild hypercapnic conditions.

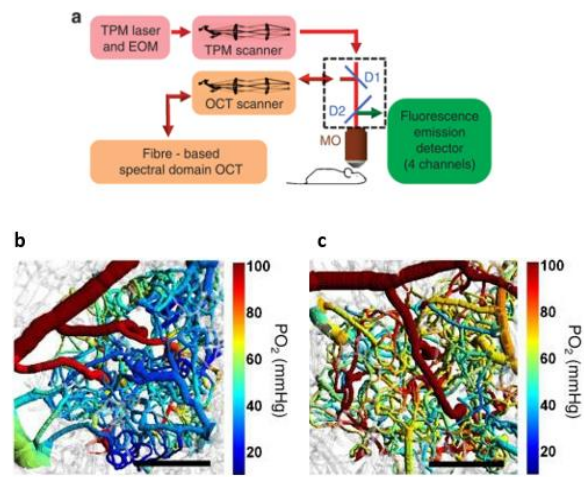

**Fig. 7** (a) Schematic of the experimental set up: Two-photon microscope (TPM), microscope objective (MO), D1 and D2-dichroic mirrors, electro-optic modulator (EOM). (b) image showing cerebral microvasculature and oxygenation during normocapnia (c) image showing cerebral microvasculature and oxygenation during hypercapnia [115] Scale bar – 200 µm.

Their results indicate that at baseline activities, arterioles extract 50% $O_2$ and the remaining $O_2$ exchange takes place at the first few capillary branches while most of the higher branching order capillaries release little $O_2$ at rest and these capillaries act as dynamic $O_2$ reserve to meet higher oxygenation demands during increased neuronal activity or decreased blood flow. Figure 7(b) and 7(c) shows images of cerebral microvasculature obtained from their experiments during normocapnia and hypercapnia respectively. Another study discusses the reconstruction of microvascular cerebral blood flow (µCBF) distributions using combined two photon microscopy and Doppler OCT [55]. In this paper, the authors investigate the reconstruction of µCBF in a truncated cortical angiogram by combining two photon microscopy and Doppler OCT in a mouse cortex down to 660 µm. In this technique the vasculature is represented as a graph consisting of nodes joined by vessel segments and the flow in each segment is computed according to Poiseuille's law. It is shown that integrating DOCT measurements along with two photon microscopy measurements greatly improves the accuracy of flow computation.

*2.1.3 Stimulated emission depletion microscopy*

Stimulated emission depletion microscopy (STED) introduced by Stefan W. Hell and Jan Wichmann in 1994 is one of several super resolution based microscopy imaging technique that overcomes the diffraction limits of microscopy enabling high spatial resolution of 20 nm and axial resolution 50 nm. STED is described briefly in [38]. It has been used to image live brain slices [24]. Apart from STED, various other super resolution microscopy techniques have been developed and used for cellular imaging studies [33].

### 2.2 Sub surface imaging techniques

### 2.2.1 Optical coherence tomography (OCT)

OCT is a 3-D tomographic imaging technique that provides high spatial and axial resolution (µm) over a depth of few millimetres (1-3 mm) for *in vivo* imaging. Similar to the above described techniques, OCT detects ballistic photons that are backscattered from various depths within the sample by utilizing coherence gating. Review of OCT theory and working principle can be found in [38]. One of the major advantages of OCT is that it provides both morphological and functional information (angiography) from the same data set. Most of these applications and techniques are described in [38]. Development of Fourier domain OCT's has enabled high speed acquisition of OCT angiograms with improved sensitivity and higher SNR and higher axial resolution [2]. In a recent paper, Choi et al. has shown imaging through a mice brain upto a depth of 2.3 mm [26].

OCT angiographic techniques utilize inter frame speckle decorrelation algorithms to detect blood flow. OCT angiographic algorithms are either amplitude based or phase based [94]. These include speckle variance, correlation mapping, optical microangiography, phase variance and Doppler OCT methods. Apart from these techniques, recently split spectrum amplitude decorrelation angiographic (SSADA) technique has been developed [73, 75] to image human macula (Figure 8) and optic nerve head which shows improved SNR to flow detection compared to other amplitude based angiographic algorithms. In this technique, in order to improve the sensitivity to flow in transverse direction and reduce sensitivity in the axial direction which is dominated by pulsatile motion in the fundus, axial resolution of the OCT voxel is broadened by creating an isotropic voxel. Attempts have been made to quantify blood flow velocity using SSADA [73, 131]. SSADA algorithm have been used in angiographic studies aimed at evaluating treatment approaches to choroidal neovascularization [128], for visualizing superficial and deep vascular networks of retinal vasculature [108, 117], in comparative studies evaluating angiographic changes in diabetic retinopathy using fluorescein angiography [71] and to study optic disc perfusion in glaucoma. SSADA has also been used to obtain quantitative parameters to detect neovascularization and in quantifying ischemia in diabetic retinopathy and in age related macular degeneration by determining flow index and vessel density [74, 131]. Speckle variance (SV) and optical microangiography (OMAG) algorithms as discussed in [38] have also been used to quantify retinal capillary vessel density with results comparable to histology [3, 4, 21, 66, 84, 147, 164]. Phase based phase variance technique has also been used to image the vasculature in the chorioretinal complex [110]. Choi et al. [28] has described a technique for imaging skin microvasculature using ultrahigh sensitive OMAG (UHS-OMAG) combined with a binary correlation mask image (mOMAG). In this technique, in order to improve the flow contrast of the angiogram, a binary mask of cmOCT mapping algorithm is applied over the UHS-OMAG image. This reduces the effect of static signals on the resulting OHS-OMAG angiogram images. mOMAG was able to detect several cutaneous layers such as stratum corneum (SC), viable epidermis (VE), papillary dermis (PD), and reticular dermis (RD) of the palmar skin. mOMAG is shown to be able to image upto 588 $\mu$m below the surface showing clearly the RD and the superficial plexus with its dense capillary network. In another study mOMAG has been used to image the microcirculation of a health human finger [11, 27, 123]. The system used a high speed swept source OCT system with centre wavelength at 1300 nm operating at 100 kHz. They were able to detect vessels upto a depth of 1380 µm showing the upper horizontal plexus that supports the dermal papillary loops and lower horizontal plexus at the dermal-subcutaneous junction. To quantify the blood flow velocity, OMAG has been combined with Doppler OCT to form DOAMG [124, 125]. RBC axial velocities ranging from 0.9 mm/s and 0.3 mm/s have been achieved by this technique. This technique has been utilized in mapping RBC velocities in capillaries of human finger cuticle [11]. OMAG has also been used to assess cutaneous wound healing process and to study acne lesion development and scarring on human facial skin [10, 155]. In another study cube correlation mapping optical coherence tomography (cube-cmOCT) have been demonstrated to map blood vessels and is shown to have improved SNR [2].

OCT angiographic techniques, both phase based and amplitude based algorithms are prone to projection artefacts [159]. These projection artefacts also called as shadow artefacts or tailing artefact causes the appearance of vessel like structures in areas which are devoid of blood flow and also elongation of blood vessels in the structural image. These artefacts arises mainly due to 1) the forward scattering of photons by RBC in the blood vessels inducing large variations in amplitude and phase of OCT signals that are collected by coherence gating along the entire depth of the tissue considered 2) changes in their path length after multiple scattering within blood vessels causing the blood vessels to appear elongated. These artefacts are more prominent in retinal angiographic images where there are high reflecting layers below the vessels as in IS/OS and RPE layers. Zhang et al. in their recent paper has suggested intensity based logarithmic scaling approach to reduce this artefact [159].

OCT has also been used to measure hemoglobin oxygen saturation (sO$_2$) in the microvasculature [29, 154]. Assessing sO$_2$ is essential for proper assessment of tissue functionality and plays an important role in ensuring adequate blood supply meeting the tissue oxygen demand. sO$_2$ measurement using OCT utilizes Fourier domain OCT. Super continuum laser is used to provide broadband illumination in the visible range and a spectrometer is used to collect interference spectrum from 520 nm to 630 nm. Unlike the other OCT angiographic techniques, in this technique wavelength-dependent B-scan images are obtained by a short-time inverse Fourier transform (STFT). To quantify sO$_2$ from oxygenated and deoxygenated Hb, angiographic images centred at four wavelengths 562.8 nm, 566.9 nm, 571.0 nm, and 575.3 nm were used based on their distinct extinction spectra [154].

There has been a growing interest in label free imaging of cerebral capillaries to detect changes in CBF. In a recent paper, Lee et al. [88] discusses a method to quantify RBC speed and flux using commercial SD-OCT on a rat cerebral cortex. The system had a centre wavelength of 1300 nm and transverse resolution of 7µm for a 5X objective. In an another study by Lee et al. [89], DLS-OCT was used to quantify CBF using the same OCT system described above but with an axial resolution of 3.5 µm. Unlike Doppler OCT, DLS–OCT is able to measure both transverse and axial flow velocities of CBF. The theory of DLS-OCT can be found in [87]. Since DLS-OCT is able to discriminate between diffusive and translational motion, it can be used to simultaneously image CBF and intracellular motility (IM). Experiments to study IM were carried out using a 40X objective lens providing a transverse resolution of 0.9 µm. Such high transverse resolution enabled the visualization of neuronal cell bodies in the minimum intensity projection image. Figure 9 (a) and 9(b) shows DLS-OCT mapping of CBF maps using a 5X objective and DLS-OCT imaging of IM using a 40X objective.

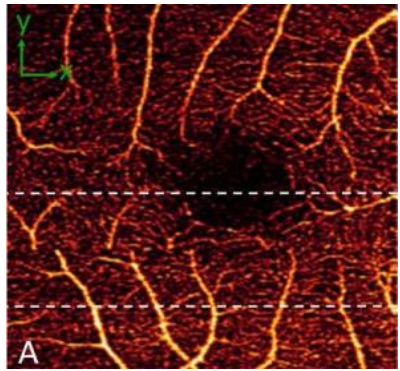

**Fig 8** *En face* OCT image of the macula showing blood vessels after SSADA algorithm [75]

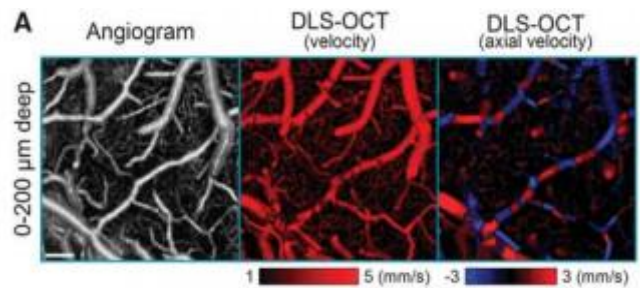

(a)

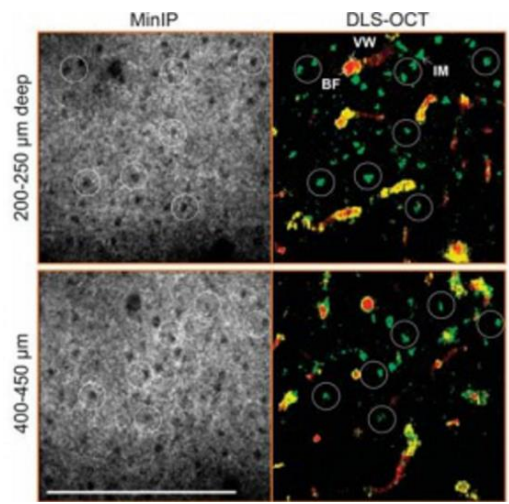

(b)

**Fig 9** (a) DLS-OCT map of CBF using 5X objective (b) DLS-OCT map of intracellular motility using 40X objective. BF- blood flow, IM- intracellular motility and VW- vessel wall. MinIP- minimum intensity projection, white corresponds to spatial correlation between the positions of neuronal cell bodies in MinIP images and neuronal IM in DLS-OCT images, Scale bar – 200 µm [89].

*2.2.2 Hyper spectral imaging*

Hyperspectral imaging (HSI) or imaging spectrometer originally introduced in remote sensing has become an emerging medical imaging technology over the past years [20, 97]. HSI is a hybrid imaging modality that combines spatial information together with spectroscopic measurements. HSI imaging technique consists of light source (visible, near-infrared or mid infra-red sources), dispersion elements (such as prism, grating) and spatial detectors (hyper spectral camera). Working principle of HSI is shown in Fig. 10. Light from the light source is used to illuminate the sample. The diffused reflected light from the sample is then passed through a collimator before getting dispersed by a grating element and focussed onto spatial detector. Thus at each spatial location, the camera registers the spectral signature at that site. Two dimensional images in HSI can be acquired either by spatial scanning or spectral scanning approach. In spatial scanning approach, 2-D images are created by scanning the sample and creating a hypercube with two spatial dimensions and one spectral dimension [20, 97]. In spectral approach, the whole sample is captured using 2-D detectors at each wavelength to form a hypercube. HSI devices mainly operate in the region of the electromagnetic spectrum called 'therapeutic window' (600 – 1300 nm), so that light can penetrate deeper into the tissues. Hypercube data over a large number of wavelengths (tens to hundreds) having spectral resolution of 2 nm can be obtained using HSI. Compared to spectroscopy which is point measurement technique, HSI provides two dimensional spectroscopic mapping which has clinical applications in microcirculation assessment and in detecting the tissue metabolic environment in tumour cells. Detailed review and application of HSI is given in [20, 97]. Combination of HSI and first pass fluorescence imaging has been used to study tumour blood flow and microvessel oxygenation in real time in nude mice [86, 100]. HSI is also used to study oxygen saturation in retinal vasculature and changes in skin microcirculation in response to varying doses of ionizing radiation [25, 49, 78, 103]. In a recent paper, an algorithm for mapping cutaneous tissue oxygen saturation has been developed [103]. In order to monitor rapid changes in vascular dynamics, snap shot multispectral imaging system has been reported recently [62]. HSI has also been combined with endoscopic system for real-time mapping of the mucosa blood supply in the lung [46]. In a recent study, polarization sensitive hyperspectral imaging was used to study the spatial distribution of melanin and hemoglobin oxygenation in skin lesions [134]. Polarisation HSI was able to separate superficial melanin so that oxy and deoxy-haemoglobin distribution from the deeper layers could be assessed.

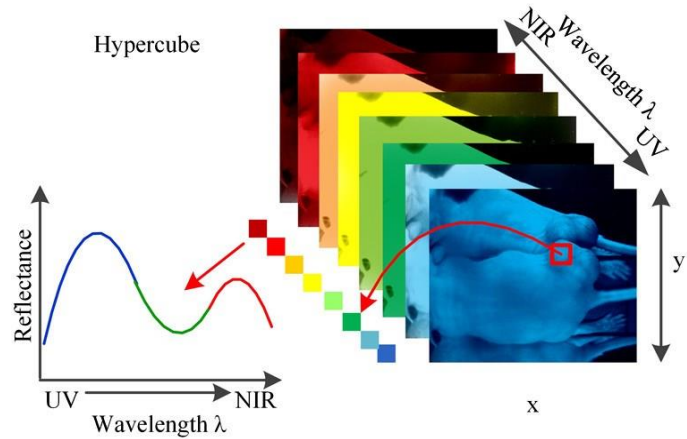

**Fig 10** Working principle of hyper spectral imaging [97].

*2.2.3 Laser speckle contrast (LSCI) and Laser Doppler perfusion imaging(LDPI)*

Laser speckle and Laser Doppler imaging have been discussed previously [38]. Recent applications of these techniques are discussed here. LSCI has been used to evaluate microcirculation in the lower extremities of ulcer patients and also to evaluate penile microvascular reactivity [102, 136]. LDPI has been used to study cutaneous microcirculation in psoriatic plaques during treatment [63]. LDPI has also demonstrated heterogeneity in perfusion within homogenous-looking plaques in psoriasis [64]. In a recent paper, relative index of blood perfusion obtained from different parts of the body using LDPI have been compared [72].

*2.2.4 Sidestream dark field (SDF) and incident dark field (IDF) imaging*

SDF techniques consists of an illumination unit used to illuminate the tissue at 530 nm, central light guide unit with lens system and an imaging module with CCD camera [56]. The illumination unit consists of concentrically placed LED's surrounding the central light guide unit. In order to overcome blurring of images due to movement of RBC's, SDF imaging uses LED's with pulsed illumination synchronized with CCD frame rate and provides stroboscopic illumination [56]. SDF provides visualization of microcirculation at the cellular level, however the maximum measurable flow velocities are limited by the frame rate of the camera (e.g. ~1 mm/s at 30 frames per second). It has been used to image nail fold capillaries, to study buccal and cutaneous microcirculation, as an adjunct to clinical free flap monitoring and to evaluate labial capillary density as a marker of coronary artery disease in diabetes [14, 36, 40, 96]. To enhance the dynamic range of RBC flow velocities, SDF has been combined with LSCI (Laser speckle contrast imaging) for in-vivo imaging [96]. Similar to SDF, another technique based on oblique profiled epi-illumination imaging has been used to study sublingual microcirculation in patients with pulmonary arterial hypertension [36]. Incident dark field imaging (IDF) which is the successor of SDF uses incident dark field imaging. Details of Cytocam-IDF imaging system is given in [8].

IDF imaging is shown to better quantify (by 30%) micro-vessels compared to SDF in sublingual mucosa [133]. Schematic of SDF and IDF imaging is shown in Figure 11(a). Figure 11(b) shows comparison between SDF and IDF images obtained from the upper arm of preterm neonates. Also IDF images have improved contrast and image sharpness. In another study evaluating the transcutaneous microcirculation in the upper arm of neonates, IDF technique was able to visualize smaller vessels compared to SDF. Also IDF was able to detect 20 % more vessels compared to SDF device [90].

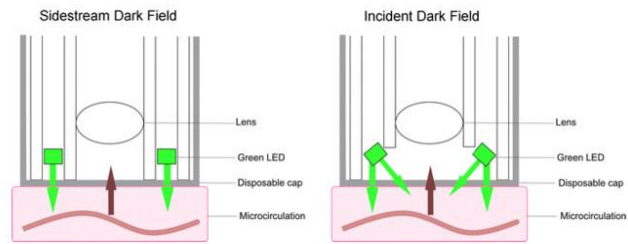

(a)

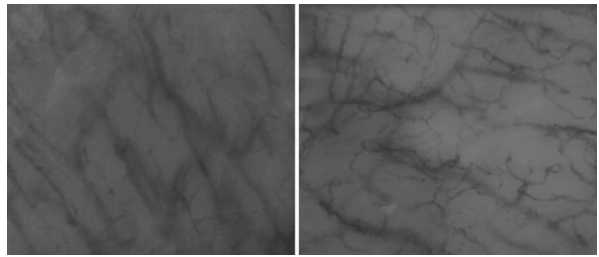

(b)

**Fig 11** (a) Schematic of SDF and IDF imaging (b) Microcirculation image of upper inner arm of preterm neonates obtained from SDF imaging (MicroScan) and IDF imaging (CytoCam) [133]

## 2.3 Deep tissue imaging techniques

### 2.3.1 Diffuse correlation spectroscopy (DCS)

Diffuse correlation spectroscopy (DCS) is a dynamic scattering method based on the temporal variations of multiple scattered near-infrared (NIR) light [42]. When light undergoes multiple scattering in a thick medium, at each scattering event, its phase is changed. Detection of these multiple scattered light fields creates a speckle pattern that varies over time. Figure 12 shows the temporal evolution of speckle intensity over time detected by a 2 D photon counting camera. As RBC's are the major scatterers within the microvasculature, the fluctuations in speckle pattern primarily indicate movement of RBC's and hence gives a measure of perfusion. DCS is a variant of dynamic light scattering (DLS) which measures the temporal autocorrelation function of single scattering photons. Similar to DLS, DCS measures the normalized intensity autocorrelation function $g_2$ given by Eq. 1 [41].

$$g_2(r,\tau) = I(r,t)I(r,t+\tau)/< I(r,t) >^2 \qquad (1)$$

where $I(r,t)$ is the intensity at time $t$ and position $r$ and $\tau$ is the correlation time. Blood flow index (BFI) is obtained by finding the mean squared displacement of the particles using a Brownian motion model. The measured BFI has the unit cm/s$^2$. As DCS uses light in NIR region, wherein tissue absorption is relatively low, photons are able to penetrate more than 1 cm deep into tissue.

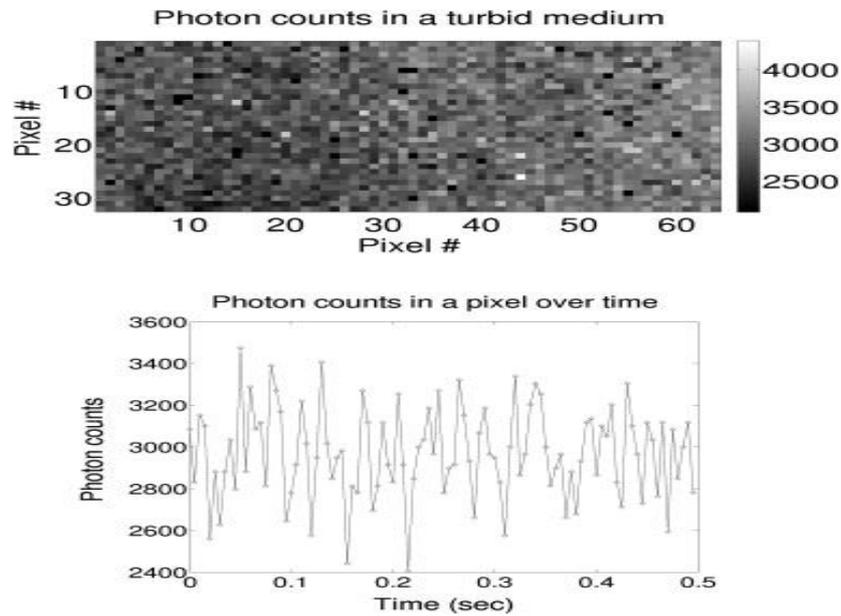

**Fig 12** Temporal fluctuation of speckle over time [41].

DCS was first used for cerebral blood flow measurements during motor stimuli in human brain through an intact skull [43]. In these studies, CBF changes measured from BFI were observed from the somatosensory cortex, visual cortex and frontal lobes in response to finger-tapping stimulus, visual stimulus and verbal fluence stimulations tasks respectively. The penetration depth of NIR light in DCS experiments depends on the source to detector separation on the tissue surface. Hence, for larger depths of penetration, the source and the detector should be placed far apart, which is not possible as the detected light decreases for large source detector separations. As DCS measures single speckle fluctuations over time, the largest optimum source detector separation achieved using a single-mode detection fibre and multimode source fibre on adult human head was found to be ~ 3 cm [42]. Parallel detection of speckle intensity fluctuations using fibre bundle and multichannel autocorrelators have been proposed by Dietsche et al. [39]. Parallel detection of intensity fluctuations have lead to improved SNR by a factor of $\sqrt{N}$ where $N$ is the number of single mode detection fibres within a fibre bundle. Numerous studies have reported the application of DCS to detect regulation of CBF in ischemic stroke patients [34, 165], patients with traumatic brain injury and sub-arachnoid damage [106]. In conjunction with near infrared spectroscopy, it has also been used to assess cerebral metabolic rate for oxygen (CMRO$_2$) of neonates with severe congenital heart defects [43] and to study the variation of in new born infants [113].

### 2.3.2 Functional near infrared spectroscopy (fNIRS)

The application of NIRS in biological tissue was first demonstrated by Jöbsis in 1977 when he studied the feasibility of detecting oxygenation changes in adult brain during hyperventilation [76]. In 1985, the first clinical studies using NIRS on newborns were reported [22]. NIRS uses light waves within the optical window of biological tissues, i.e. 650-1400 nm. Within this window, biological tissue is relatively transparent and absorption by haemoglobin (Hb), oxy-haemoglobin and deoxy-haemoglobin dominates over scattering. The basic principle of fNIRS is that the differences in the absorption spectra of oxy-Hb and deoxy-Hb at two or more wavelengths within the spectral window can be used to measure the relative changes in oxy-Hb concentration. This effect has been exploited in pulse oximeters. The first fNIRS measurements were reported independently by groups Chance et al. and Villringer et al. in 1992 [22, 136]. The development and history of fNIRS spectroscopy can be found in [48, 120, 138]. Currently, there are three different variants of NIRS, namely continuous wave spectroscopy (CWS), time of flight spectroscopy and frequency domain spectroscopy [138]. CWS uses an NIR light source to illuminate the sample and the change in intensities of the diffusely reflected light collected from the sample is analysed. Since, its introduction CWS has been used to study cerebral haemodynamics due to neurovascular coupling. Together with fMRI, fNIRS has been used to provide cerebral blood oxygenation changes during stimulation of the human brain. The first such study described the changes in blood oxygenation in the cortical region due to functional activation of sensory motor cortex during a unilateral finger opposition task [81].

In fNIRS, the intensity changes of the backscattered light from the sample are quantified to detect the chromophore concentrations using modified Beer-Lambert law given by Eq. 2 that gives the relationship between the extent of light attenuation within tissue and the concentrations of the chromophores (*c* in *mM*) [17, 79]. If tissue chromophores considered are oxy - Hb and deoxy - Hb, then according to modified Beer-Lambert law,

$$-\log\left(\frac{I(\lambda,t)}{I_0(\lambda,t)}\right) = \epsilon_{oxy}(\lambda)C_{oxy}(t)d + \epsilon_{deoxy}(\lambda)C_{deoxy}(t)d + a(\lambda,t) \quad (2)$$

where $I(\lambda,t)$ is the incident light of wavelength λ at time $t$, $I_0(\lambda,t)$ is the diffusely reflected light, $\epsilon$ is the molar absorbances of each of the chromophores (*mm/ mM*), d is the optical path (*mm*) that the light travels within the tissue and '*a*' is the term that includes the absorption and scattering effects due to other chromophores within the probed tissue. From Eq. 2, the changes in total Hb content can be found. The measured changes in oxy- Hb and deoxy- Hb are dependent on the path length of light travelled in the tissue. This causes significant problem for comparison across subjects and also the path length varies depending on factors such as source to detector separation and underlying tissue morphology. This is a limiting factor of CWS while frequency domain and time of flight spectroscopy variants of NIRS allows the determination of path lengths. CWS is preferred over frequency domain and time of flight spectroscopy as these techniques are much sophisticated and expensive. Another method to enable path length independent measurement is to use dual wavelength spectroscopic measurements. There has been a lot of research into selecting the optimum wavelengths based on theoretical and experimental approaches. Yamashita et al. [149] showed that wavelength pair 664 nm and 830 nm are optimum for oxy- deoxy -Hb measurement than 780 nm and 830 nm. Funane et al. [54] from his theoretical studies reported that maximum SNR is obtained when both ends of the range of 659–900 nm were used. Correia et al. [32] reported that optimum wavelength pair to be used is 704 ± 7 and 887 ± 12 nm. There have also been studies using three and four wavelength sources by considering lipid and water as additional chromophores [31, 166] and it was found that 782, 832 and 884 nm are optimal for three wavelength sources, and that 786, 807, 850 and 889 nm are optimal for four wavelength sources. Numerous algorithms have been developed to improve the signal processing of the acquired fNIRS spectra and also to separate them into neuronal and systemic parts using mathematical modelling [68, 119, 120, 148, 156, 158].

The applications of fNIRS to study cerebral haemodynamics have been reviewed in the past and has been mainly used to study neurovascular coupling and rate of oxygen consumption in the brain ($CMRO_2$) and in brain computer interfacing [129]. In a study conducted by Franceschini et al. on evoked somatosensory potentials [52] combining diffuse optical imaging and EEG, it has been shown that evoked haemoglobin response is driven by the cortical synaptic activity and not by direct thalamic input. To measure $CMRO_2$ indiectly from fNIRS, Windkessel model was used to relate changes in dynamic blood volume to the changes in cerebral blood flow using wavelength 682 nm and 830 nm [16]. To reduce motion artefacts and improve the optode-scalp coupling, miniaturized optical fibre probes with tips fixed to the scalp using collodion have recently been developed [156]. fNIRS have been used to monitor CBF changes during epileptic seizures [1, 126, 143]. In a recent paper, Roche-Labarbe et al. [114] measured oxy- and deoxy-hemoglobin concentrations, CBV, relative CBF (rCBF) and relative $CMRO_2$ ($rCMRO_2$) changes in the somatosensory cortex of preterm neonates during passive tactile stimulation of the hand by combining DCS, frequency-domain near-infrared and CWS. This study is the first to report local rCBF and $rCMRO_2$ during functional activation in preterm infants. They observed that blood flow increased immediately after the onset of stimulus and returned to baseline prior to Hb concentration, oxygenation, and blood volume suggesting higher compliance of veins relative to arterioles. Recently fNIRS has been used to evaluate haemodynamic response to innocuous and a noxious electrical stimulus on healthy human subjects and discusses its potential utility to measure pain [157].

*2.3.3 Photoacoustic tomography and microscopy*

In photoacoustic imaging (PAI), light pulses are used to generate ultrasound waves within the tissue produced by the optical absorption of chromophores. In PAI, nano second pulses (NIR) are delivered onto the tissue of interest, where in, it is locally absorbed by tissue chromophores and is converted to heat. This generated heat causes the generation of pressure waves by thermoelastic expansion and the pressure waves propagate as ultrasound/ photoacoustic waves which can be detected by an ultrasound transducer. As PAI uses NIR light, it can penetrate several centimetres within the tissue and provide map of vasculature from the haemoglobin oxygen saturation in blood. Moreover, PAI generates high contrast vasculature map as it is based on intrinsic tissue chromophore absorption and is free from speckles that degrade image quality in OCT and ultrasound imaging modalities. Briefly the theory and working of PAI has been reviewed in [38]. PAI has been widely used in vascular imaging. PAI is mainly of 2 types: Photoacoustic tomography (PAT) and photoacoustic microscopy (PAM) [132, 146].

PAT provides spatial resolution of 0.1mm to 1 mm and an axial resolution of 1 mm depending on the bandwidth of the ultrasound detector used and the light pulse duration respectively. PAT is shown to be capable of imaging upto 5 mm in biological tissue [132]. PAT generally uses low frequency ultrasound transducers < 10 MHz to image depths greater than 1 cm.
Although high frequency ultrasound transducer probes offer finer lateral resolution, they provide shallower imaging depths as they are attenuated faster compared to low frequency probes. Apart from visualizing vasculature, PAT has also been used to quantify blood flow. To quantify blood flow, Doppler shift of the photoacoustic modulation has been explored similar to spectral ultrasound Doppler imaging. Photoacoustic Doppler imaging is shown to detect flow velocities from 0.014 to 3700 mm/s [142, 151, 162]. Recently, photoacoustics lifetime imaging (PALI) was developed to map spatial and temporal distribution of tissue oxygen by measuring the triplet state lifetime of a chromophore (oxygen sensitive dye) [121,122]. PAI has been used to study cerebral vasculature [82] and recently five dimensional photoacoustic tomography providing 3-D multispectral imaging in real time have been developed to study cerebral haemodynamics in mouse brains [57]. PAT has also been used to study neurovascular coupling [139]. Intravascular photoacoustic tomography has been used to study lipid deposition [67, 161]. PAI is also studied in early detection and diagnosis of tumour and for monitoring treatments effects over time [77, 85, 98, 111]. Exogenous contrast agents such as gold nano particles that provide optical contrast by absorption of light have been used to image tumour vasculature and neovascularization [18, 44, 91, 105].

Compared to PAT, PAM is able to achieve superior spatial resolution of 0.2-60 µm and imaging depth of several mm [151]. PAM is further divided into OR-PAM (optical resolution PAM) and AR-PAM (acoustic resolution PAM) based on tight optical focussing and acoustic focussing respectively. OR-PAM provides high spatial resolution of 1-2 µm but has limited imaging depth of ~1 mm in muscle and ~0.6 mm in the brain due to the optical scattering in turbid media [146] whereas the AR-PAM has an imaging depth of several millimetres but has spatial resolution of 50-60 µm. By using diffraction limited focussing PAM (sub wavelength PAM), lateral resolution of 220 nm was achieved for an imaging depth of 100 µm [160]. Flow velocities in the range of 0.01 to 25 mm/s have been measured using structural illumination based PAM [163] and temporal correlation techniques [92, 127]. Similar to PAT, PAM has been used to for quantitative microvascular imaging [150, 162], tumour angiogenesis [93] cerebral imaging in mouse (imaging depth of 0.7 µm at 250 µm beneath the skull) [153] and in imaging lymph vessels [99]. Another technique called PA flowoxigraphy (FOG) based on high resolution PAM have been developed which can detect dynamic oxygen release from single flowing RBC [141]. FOG has micrometer spatial resolution and oxygen detection time of 20 µs. FOG was also shown to be capable of simultaneously quantifying RBC functional parameters such as total haemoglobin concentration, oxygen saturation (sO2), sO2 gradient, flow speed, and oxygen release rate [141]. Representative image of sO2 map using FOG is shown in figure 13.

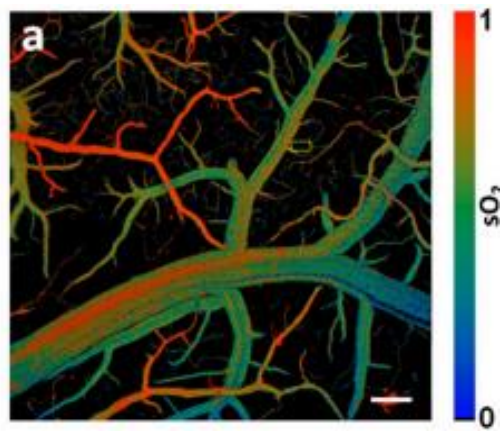

**Fig 13** sO2 map of the mouse brain cortex obtained from photoacoustic flowoxiography [141]

## 3. Recent applications of optical imaging in cerebral haemodynamics

In this section, we describe the applications of optical imaging techniques that have been used in recent years to gain better understanding of cerebral haemodynamics.

*3.1 Haemodynamic point spread function*

Some authors [83] argue that resolution better than 2 mm is unhelpful in the study of neurocoupling, the impact of neural activity on the microcirculation, since the impact of even the smallest stimulated focal volume impacts the microcirculation on a scale of approximately 2.3 mm [45].

### 3.2 Role of pericytes in cerebral blood flow regulation

Pericytes were first observed by Eberth and Rouget in 1870 as spatially isolated cells on the outer walls of capillaries and at the capillary branch points [7]. These cells were thought to have contractile properties thereby controlling vessel diameter, vascular resistance and thus playing an important role in the cerebral blood flow regulation (CBF). Apart from the contractile property, pericytes were also thought to play an important role in angiogenesis, maintenance of blood brain barrier and in microvascular development [7, 37]. However, there has been much debate over whether it's the pericytes that regulate CBF by its contractility or is it the smooth muscle cells (SMC) that line the arterioles and venules. Recent studies have used two photon microscopic imaging to tackle this problem. In experiments conducted using transgenic mice labelled with fluorescent proteins (NG2 and α SMA promoters to label the mural cells) Hill et al. [65] observed that capillary pericytes lack actin and hence are not contractile in vivo. By their targeted neuronal stimulation using optogenetics, whisker stimulation and cortical spreading depolarization, they report that changes in microvascular diameter and flow occur in microvessels covered by SMC's but not in those covered by pericytes. They also reported that during early brain ischemia, it is the transient SMC and not the pericytes that cause hypoperfusion. This is in contradiction to the results published by Hall et al. [60] who demonstrated that pericytes are the first vascular element to dilate during neuronal activity in the neocortex and cerebellum and that they die during ischemia. By combing two-photon imaging and optogenetics for imaging the cerebral vasculature they observed that during whisker stimulation, $1^{st}$ order capillaries dilated before penetrating arterioles and also the change in capillary diameter (by over 5%) was more in regions with pericytes than in pericyte free zones. These findings support the fact that most neurons being closer to capillaries than to the arterioles regulate cerebral microcirculation by signalling to the pericytes thus dilating the capillaries before arterioles. However, these findings contradict the results reported by Fernández-Kletta et al. [47] who reported that precapillary and penetrating arterioles play a major role in increased blood flow induced by neural activity than capillary pericytes by using intravital two-photon laser scanning microscopy and exposing the pericytes to potent vasoconstrictors. The differences in these findings can be attributed to variations in induced neuronal stimulation techniques which may have triggered the release of vasoconstricting messengers, the use of different anaesthetic agent that may have suppressed the blood flow increases and finally could be due to defining $1^{st}$ order capillaries as precapillary arterioles due to their morphological similarities [65]. A recent paper by Attwell et al. [7] suggests that these contradicting results occur mainly due to the differences in nomenclature of pericytes among these papers. Two photon imaging has been used to characterize and label pericytes in the cerebral cortex [61]. Figure 14(a) shows the schematic of different types of pericyte morphologies. and a two photon image. Figure 14(b) and 14(c) show two photon images of the capillary bed obtained from the cortex of a transgenic mouse indicating capillary pericytes within them.

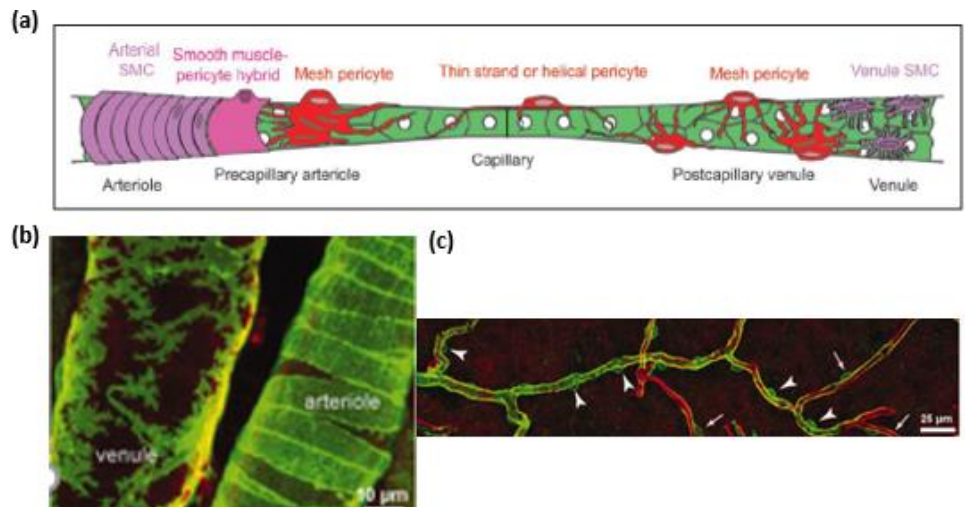

**Fig 14** (a) Schematic of different types of pericyte morphology in capillary bed [61] (b) Two photon image from the cortex of an NG2cre:mT/mG transgenic mouse showing mGFP (green) expressing smooth muscle cells on arterioles and venules with mTomato (red) expression in all membranes. (c) Images showing transitions from mGFP-labeled smooth muscle cell morphology to capillary pericytes , mGFP labelled smooth muscle cells denoted by arrowheads and capillary pericytes denoted by arrows [65].

Two photon microcirculation imaging with optogenetics is a hot topic which will finally resolve these conflicts.

## 4. Conclusion

This review has provided an update of currently available microcirculation imaging techniques that has potential clinical applications. The review briefly discusses the various technological advances behind various microcirculation imaging techniques, its resolution limit, penetration depth and applications. Most of the discussed techniques are able to detect morphological and functional information of the underlying biological tissue. Also, by combining different imaging modalities additional information can be obtained which otherwise wouldn't be possible.